\documentclass[twocolumn,showpacs,preprintnumbers,amsmath,amssymb]{revtex4}

\usepackage{graphicx}
\usepackage{dcolumn}
\usepackage{bm}

\begin{document}

\title{Nonequilibrium Extension of the Landau-Lifshitz-Gilbert Equation
for Magnetic Systems}

\author{Jeongwon Ho}
\affiliation{Department of Physics and Astronomy, University of
Victoria, Victoria, British Columbia, Canada}
\author{B. C. Choi}
\affiliation{Department of Physics and Astronomy, University of
Victoria, Victoria, British Columbia, Canada}
\author{F. C. Khanna\footnote{Permanent address: Department of Physics, University of Alberta,
Edmonton, Alberta, Canada} and Sang Pyo Kim}
\affiliation{Department of Physics, Kunsan National University,
Kunsan 573-701, Korea}

\begin{abstract}
Using the invariant operator method for an effective Hamiltonian
including the radiation-spin interaction, we describe the quantum
theory for magnetization dynamics when the spin system evolves
nonadiabatically and out of equilibrium, $d \hat{\rho}/dt \neq 0$.
It is shown that the vector parameter of the invariant operator
and the magnetization defined with respect to the density
operator, both satisfying the quantum Liouville equation, still
obey the Landau-Lifshitz-Gilbert equation.
\end{abstract}

\maketitle

\section{Introduction}

Magnetization dynamics has attracted much attention both in
theoretical and experimental physics. In particular, the magnetic
information requires a deep and fundamental understanding of the
dynamics of spin system on a short time scale. The
Landau-Lifshitz-Gilbert (LLG) equation provides a plausible
phenomenological model for many experimental results \cite{llg}.
Recently, the first three authors have derived the LLG equation
from an effective Hamiltonian including the radiation-spin
interaction \cite{hkc}. It is assumed there that the magnetic
system maintains quasi-adiabatic evolution and obeys the condition
$d \hat{\rho}/dt = 0$.

However, for a magnetic system whose Hamiltonian $\hat{\cal H}
(t)$ evolves nonadiabatically, its statistics deviates far from
quasi-equilibrium.  Thus the density operator does neither satisfy
the condition $d \hat{\rho}/dt = 0$ nor is given by $e^{- \beta
\hat{\cal H} (t)}$. Instead, the density operator {\it does}
satisfy the quantum Liouville equation
\begin{equation}
i \hbar \frac{\partial \hat{\rho}}{\partial t} + [ \hat{\rho},
\hat{\cal H}] = 0. \label{leq}
\end{equation}
The main purpose of this paper is to derive the LLG equation from
quantum theory for such a nonequilibrium magnetic system with the
effective Hamiltonian in Ref. \cite{hkc}.

To find the nonadiabatic quantum states for this nonequilibrium
system, we employ the invariant method developed for explicitly
time-dependent Hamiltonians in the Schr\"{o}dinger picture
\cite{lewis}. One advantage of the invariant operator method is
that the eigenstates of the invariant operator satisfying the
quantum Liouville equation are exact quantum states of the
time-dependent Schr\"{o}dinger equation up to time-dependent phase
factors. Further we use this invariant operator to define the
density operator \cite{kim}, in terms of which we define the
magnetization. Finally we show that the magnetization satisfies
the LLG equation even for nonadiabatic and nonequilibrium
evolution.

\section{Landau-Lifshitz Equation}

We first consider the spin Hamiltonian without the radiation-spin
interaction given by
\begin{equation}
\hat{\cal H}_0 (t) = - g \mu_B \sum_{i = 1}^{N} \hat{S}_i \cdot
{\bf H}_{eff} (t), \label{ham0}
\end{equation}
where $g$ is the Lande's g-factor, $\mu_B$ is the Bohr magneton,
$N$ is the number of spins in the system, and $\hat{S}_i$ is the
$i$th spin operator. The effective field ${\bf H}_{eff}$ includes
the exchange field, the anisotropy field, and the demagnetizing
field as well as the external field. As the Hamiltonian
(\ref{ham0}) has the group $SU^N(2)$, we may look for an invariant
operator with the same group $SU^N(2)$ of the form \cite{mizrahi}
\begin{equation}
\hat{\cal I}_0 (t) = \sum_{i = 1}^{N} \hat{S}_i \cdot {\bf R}_0
(t), \label{inv op0}
\end{equation}
where ${\bf R}_0$ is a vector parameter to be determined by a
dynamical equation. The invariant operator, satisfying the quantum
Liouville equation (\ref{leq}), leads to the equation
\begin{equation}
- i \hbar \sum_{i = 1}^{N} \hat{S}_i \cdot \Biggl( \frac{d {\bf
R}_0}{dt} + g \mu_B {\bf R}_0 \times  {\bf H}_{eff} \Biggr) = 0.
\end{equation}
We thus obtain the equation for the vector parameter:
\begin{equation}
\frac{d {\bf R}_0}{dt} = - \gamma {\bf R}_0 \times  {\bf H}_{eff}
\end{equation}
with $\gamma = g \mu_B$.

We note that the eigenstates of the invariant operator (\ref{inv
op0}) are exact quantum states, up to time-dependent phase
factors, of the time-dependent Schr\"{o}dinger equation for the
Hamiltonian (\ref{ham0}). Hence the vector parameter ${\bf R}_0$
defines the magnetization ${\bf R}_0 = {\bf M}_0$ of the system
during the nonadiabatic evolution. When the Hamiltonian $\hat{\cal
H} (t)$ explicitly depends on time, not only the states but also
the operators change in the invariant method.  So any physical
quantity defined by
\begin{equation}
{\bf O}(t) = {\bf Tr} \{\hat{\rho} (t) \hat{\cal O} (t)\}
\end{equation}
has, in general, the time derivative
\begin{eqnarray}
\frac{d {\bf O}(t)}{dt} &=&  \sum_{n} \langle \psi_n (t) \vert
\frac{1}{i \hbar}[\hat{\rho} \hat{\cal O}, \hat{\cal H}] +
\frac{\partial \hat{\rho}}{\partial t}
 \hat{\cal O} + \hat{\rho} \frac{\partial \hat{O}}{\partial t}
\vert \psi_n (t) \rangle \nonumber\\
&=& {\bf Tr} \Bigl\{\frac{1}{i \hbar} \hat{\rho} [\hat{\cal O},
\hat{\cal H}] + \hat{\rho} \frac{\partial \hat{O}}{\partial t}
\Bigr\}, \label{gen eq}
\end{eqnarray}
where we use Eq. (\ref{leq}) in the last step.

In the first case of no radiation-spin interaction, the density
operator may be given by \cite{kim}
\begin{equation}
\hat{\rho}_0 (t) = \frac{1}{\cal Z}_0 e^{ - \beta \hat{\cal I}_0
(t)}, \quad {\cal Z}_0 = {\bf Tr} \{ e^{ - \beta \hat{\cal I}_0
(t)} \},
\end{equation}
where $\hat{\cal I}_0$ already satisfied Eq. (\ref{leq}). Here
$\beta$ is the inverse temperature. Now we may define the
magnetization per volume in the general case as
\begin{equation}
{\bf M}_0 (t) = \frac{1}{V} {\bf Tr} \{ \hat{\rho}_0 (t) \hat{\cal
M}_0 \}, \label{m0}
\end{equation}
where $V$ is the volume of the system and ${\cal M}_0$ is the
magnetic moment operator defined by the external field as
\begin{equation}
\hat{\cal M}_0 = - \frac{\delta {\cal H}_0}{\delta {\bf H}_{eff}}
= g \mu_B \sum_{i = 1}^{N} \hat{S}_i.
\end{equation}
Then it follows from Eq. (\ref{gen eq}) that
\begin{equation}
\frac{d {\bf M}_0}{dt} = - \gamma {\bf M}_0 \times  {\bf H}_{eff}.
\label{llg0}
\end{equation}
Note that Eq. (\ref{llg0}) is the Landau-Lifshitz equation.

\section{Landau-Lifshitz-Gilbert Equation}

In Ref. \cite{hkc} the LLG equation is described by an effective
theory including the radiation-spin interaction. The model
Hamiltonian
\begin{equation}
\hat{\cal H} (t) = \hat{\cal H}_0 (t) + \lambda \hat{\cal H}_{int}
(t) ,
\end{equation}
with the radiation-spin interaction
\begin{equation}
\hat{\cal H}_{int} = g \mu_B \sum_{i = 1}^{N} \hat{S}_i \cdot
\Bigl(\alpha M^2 {\bf H}_{eff} - {\bf M} \times {\bf H}_{eff}
\Bigr), \label{rsi}
\end{equation}
is an effective theory in the sense that the Hamiltonian involves
the magnetization to be determined by the theory itself. As
$\hat{\cal H}_{int}$ has the same group structure $SU^N(2)$ as
$\hat{\cal H}_0$, we still have an invariant operator of the same
form
\begin{equation}
\hat{\cal I} (t) = \sum_{i =1}^{N} \hat{S}_i \cdot {\bf R} (t).
\label{inv op}
\end{equation}
Then the quantum Liouville equation (\ref{leq}) leads to the
parameter vector equation
\begin{equation}
\frac{d {\bf R}}{dt} = - g \mu_B {\bf R} \times \Bigl( (1 -
\lambda \alpha M^2)  {\bf H}_{eff} + \lambda {\bf M} \times {\bf
H}_{eff} \Bigr). \label{pa eq}
\end{equation}
Noting again that the invariant operator (\ref{inv op}) determines
exact eigenstates, we may identify ${\bf R} = {\bf M}$ for the
nonadiabatic evolution. Using ${\bf M} \cdot d{\bf M}/dt = 0$ and
solving for ${\bf M} \times d{\bf M}/dt$, we obtain
\begin{equation}
\frac{d {\bf M}}{dt} = - \gamma {\bf M} \times {\bf H}_{eff} +
\alpha {\bf M} \times \frac{d{\bf M}}{dt}.
\end{equation}
Thus the spin direction of the system even for the nonadiabatic
evolution obeys the LLG equation.

In the nonequilibrium case, using the density operator
\begin{equation}
\hat{\rho} (t) = \frac{1}{\cal Z} e^{ - \beta \hat{\cal I} (t)},
\quad {\cal Z} = {\bf Tr} \{ e^{ - \beta \hat{\cal I} (t)} \},
\label{den op}
\end{equation}
we may define the magnetization as
\begin{equation}
{\bf M} (t) = \frac{1}{V} {\bf Tr} \{ \hat{\rho} (t) \hat{\cal M}
(t) \},
\end{equation}
where ${\cal M} (t)$ is the magnetic moment operator
\begin{eqnarray}
\hat{\cal M} = - \frac{\delta {\cal H}}{\delta {\bf H}_{eff}} = g
\mu_B \sum_{i =1}^{N} \Bigl\{ (1 - \lambda \alpha M^2) \hat{S}_i +
\lambda \hat{S}_i \times {\bf M} \Bigr\}.
\end{eqnarray}
The magnetization can be written as
\begin{equation}
{\bf M} = (1 - \lambda \alpha M^2) {\bf M}_1 + \lambda {\bf M}_1
\times {\bf M}, \label{eq 18}
\end{equation}
where
\begin{equation}
{\bf M}_1 (t) = \frac{1}{V} {\bf Tr} \Bigl\{\hat{\rho} \hat{\cal
M}_0 \Bigr\}. \label{m1}
\end{equation}
It should be noted that ${\bf M}_0$ and ${\bf M}_1$ are weighted
with $\hat{\rho}_0$ and $\hat{\rho}$, respectively. Taking the
cross product of Eq. ({\ref{eq 18}) with ${\bf M}_1$ and
rearranging ${\bf M}_1 \times ({\bf M}_1 \times {\bf M})$, we find
that ${\bf M}$ is parallel to ${\bf M}_1$:
\begin{equation}
{\bf M} = \frac{(1 - \lambda \alpha M^2) + \lambda^2 ({\bf M}_1
\cdot {\bf M})}{1 + \lambda M_1^2}~~ {\bf M}_1.
\end{equation}
Hence the last term in Eq. (\ref{eq 18}) drops out, and we get
\begin{equation}
{\bf M} = (1 - \lambda \alpha M^2) {\bf M}_1. \label{eq 21}
\end{equation}
The magnetic moment may depend on time through ${\bf M}$. From Eq.
(\ref{gen eq}) follows the time derivative of ${\bf M}$, which is
given by
\begin{eqnarray}
\frac{d {\bf M}}{dt} = \frac{i}{\hbar V} {\bf Tr} \{ \hat{\rho}
[\hat{\cal H}, \hat{\cal M} ] \} + \frac{1}{V} {\bf Tr} \Bigl\{
\hat{\rho} \frac{\partial \hat{\cal M}}{\partial t} \Bigr\}.
\label{master eq}
\end{eqnarray}
Note that Eq. (\ref{master eq}) is quite a general result, valid
in all circumstances, including nonequilibrium evolution.
Furthermore, it has the same form as Eq. (15) of Ref. \cite{hkc},
where the static limit $d \hat{\rho}/dt = 0$ was assumed.
Evaluating Eq. (\ref{master eq}), we finally obtain the LLG
equation
\begin{eqnarray}
\frac{d {\bf M}}{dt} = - \gamma {\bf M} \times {\bf H}_{eff} +
\alpha {\bf M} \times \frac{d{\bf M}}{dt}. \label{master eq2}
\end{eqnarray}
In deriving Eq. (\ref{master eq2}) from Eq. (\ref{master eq}) we
have used ${\bf M} \cdot d {\bf M}/dt = 0$, which can be checked
self-consistently in Eq. (\ref{master eq2}).

A few comments are in order. The Gilbert damping term in Eq.
(\ref{master eq2}) has the effect of slowing down the precession
of magnetization and leads to a final constant value. Thus the
final state is a thermal state with the constant magnetization.
Then the vector parameter in Eq. (\ref{pa eq}) would have the
final value
\begin{equation}
{\bf R} (t = \infty) = - g \mu_B \Bigl( (1 - \lambda \alpha M^2)
{\bf H}_{eff} + \lambda {\bf M} \times {\bf H}_{eff} \Bigr).
\end{equation}
The invariant operator settles down to the Hamiltonian itself,
$\hat{\cal I} (\infty) = \hat{\cal H} (\infty)$. Thus the system,
after undergoing a nonequilibrium evolution, reaches a
thermalization process towards the final equilibrium with
$\hat{\rho} (\infty) = e^{ - \beta \hat{\cal H} (\infty)}/{\cal
Z}$.

\section{Conclusion}

In this paper we derived the Landau-Lifshitz-Gilbert equation for
the magnetic system with an effective Hamiltonian including the
radiation-spin interaction in Ref. \cite{hkc}. The time-evolving
magnetization enters the Hamiltonian as a parameter and the
Hamiltonian thus provides an effective theory. When the
magnetization proceeds rapidly, the system evolves
nonadiabatically and out of equilibrium. To treat such a
nonequilibrium evolution, we employed the invariant method to find
the equation for magnetization.

The invariant operator, satisfying the quantum Liouville equation,
provides exact quantum states, up to time-dependent phase factors,
of the time-dependent Schr\"{o}dinger equation. Due to the group
structure of the effective Hamiltonian (\ref{rsi}), we were able
to find the invariant operator, $\hat{\cal I} (t)$ in Eq.
(\ref{inv op}), whose vector parameter defines the magnetization.
We further used the invariant operator to introduce the density
operator (\ref{den op}), in terms of which the magnetization,
${\bf M} (t) = {\bf Tr} \{ e^{- \beta \hat{\cal I} (t) \hat{\cal
M} (t)} \}/{\cal Z}$, is defined for nonequilibrium evolution. We
showed that the dynamical equation for nonequilibrium
magnetization satisfies the same equation as for the equilibrium
case and, therefore, the Landau-Lifshitz-Gilbert equation is valid
in all cases of time development. The nonequilibrium definition of
the magnetization in this paper has the following physical
meaning. After the magnetization reaches a final value, the
invariant operator reduces to the Hamiltonian itself, $\hat{\cal
I} (\infty) = \hat{\cal H} (\infty)$, and the magnetization with
respect to the density operator, $e^{- \beta \hat{\cal I}}$, is
nothing but a thermal ensemble average of the magnetic moment
operator.

\acknowledgments

F.C.K would like to appreciate the warm hospitality of Kunsan
National University. The work of J.H., F.C.K and B.C.C was
supported in part by the Natural Sciences and Engineering Research
Council of Canada and the work of S.P.K by the Korea Research
Foundation under Grant No. KRF-2003-041-C20053.

\end{document}